# Pion Polarizability Status Report

## Murray Moinester

School of Physics and Astronomy
Tel Aviv University, 69978 Tel Aviv, Israel
E-mail: murray.moinester@gmail.com
http://www-nuclear.tau.ac.il/~murraym

**Abstract**

The electric $\alpha_\pi$ and magnetic $\beta_\pi$ charged pion Compton polarizabilities are of fundamental interest in the low-energy sector of quantum chromodynamics (QCD). They are directly linked to the phenomenon of spontaneously broken chiral symmetry within QCD and to the chiral QCD lagrangian. The combination $(\alpha_\pi - \beta_\pi)$ was measured by: (1) CERN COMPASS via radiative pion Primakoff scattering (Bremsstrahlung) in the nuclear Coulomb field, $\pi^- Z \to \pi^- Z \gamma$, (2) SLAC PEP Mark-II via two-photon production of pion pairs, $\gamma\gamma \to \pi^+\pi^-$, and (3) Mainz Microtron MAMI via radiative pion photoproduction from the proton, $\gamma p \to \gamma \pi^+ n$. COMPASS and Mark-II agree with one another: (1) $\alpha_\pi - \beta_\pi = (4.0 \pm 1.2_{stat} \pm 1.4_{syst}) \times 10^{-4} fm^3$, (2) $\alpha_\pi - \beta_\pi = (4.4 \pm 3.2_{stat+syst}) \times 10^{-4} fm^3$. The Mainz value
(3) $\alpha_\pi - \beta_\pi = (11.6 \pm 1.5_{stat} \pm 3.0_{syst} \pm 0.5_{model}) \times 10^{-4} fm^3$ is excluded on the basis of a dispersion relations calculation which uses the Mainz value as input, and gives significantly too large $\gamma\gamma \to \pi^0\pi^0$ cross sections compared to DESY Crystal Ball data. COMPASS and Mark-II polarizability values agree well with the two-loop chiral perturbation theory (ChPT) prediction $\alpha_\pi - \beta_\pi = (5.7 \pm 1.0) \times 10^{-4} fm^3$, thereby strengthening the identification of the pion with the Goldstone boson of QCD.

## 1. Introduction

The electric $\alpha_\pi$ and magnetic $\beta_\pi$ charged pion polarizabilities characterize the induced dipole moments of the pion during $\gamma\pi$ Compton scattering. These moments are induced via the interaction of the $\gamma$'s electromagnetic field with the quark substructure of the pion. In particular, $\alpha_\pi$ is the proportionality constant between the $\gamma$'s electric field and the electric dipole moment, while $\beta_\pi$ is similarly related to the $\gamma$'s magnetic field and the induced magnetic dipole moment [1]. The polarizabilities are fundamental characteristics of the pion. A stringent test of chiral perturbation theory (ChPT) is possible by comparing the experimental polarizabilities with the chiral perturbation theory ChPT two-loop predictions $\alpha_\pi - \beta_\pi = (5.7 \pm 1.0) \times 10^{-4} fm^3$ and $\alpha_\pi + \beta_\pi = 0.16 \times 10^{-4} fm^3$ [2]. The pion polarizability combination $(\alpha_\pi - \beta_\pi)$ may be measured by four different methods, as illustrated in Fig. 1. These are (1) radiative pion Primakoff scattering (pion Bremsstrahlung) in the nuclear Coulomb field $\pi Z \to \pi Z \gamma$, (2) two-photon fusion production of pion pairs $\gamma\gamma \to \pi\pi$ via the $e^+e^- \to e^+e^-\pi^+\pi^-$ reaction, (3) radiative pion photoproduction from the proton $\gamma p \to \gamma \pi n$, and (4) Primakoff scattering of high energy $\gamma$'s in the nuclear Coulomb field leading to two-photon fusion production of pion pairs $\gamma\gamma \to \pi\pi$. Methods 1,2,3 experiments have been most recently studied at CERN COMPASS [3], SLAC PEP Mark-II [4], and Mainz Microtron MAMI [5] respectively; while the method 4 experiment is planned at Jefferson Laboratory (JLab) [6]. The measured values are:
(1) $\alpha_\pi - \beta_\pi = (4.0 \pm 1.2_{stat} \pm 1.4_{syst}) \times 10^{-4} fm^3$, (2) $\alpha_\pi - \beta_\pi = (4.4 \pm 3.2_{stat+syst}) \times 10^{-4} fm^3$,
(3) $\alpha_\pi - \beta_\pi = (11.6 \pm 1.5_{stat} \pm 3.0_{syst} \pm 0.5_{model}) \times 10^{-4} fm^3$.

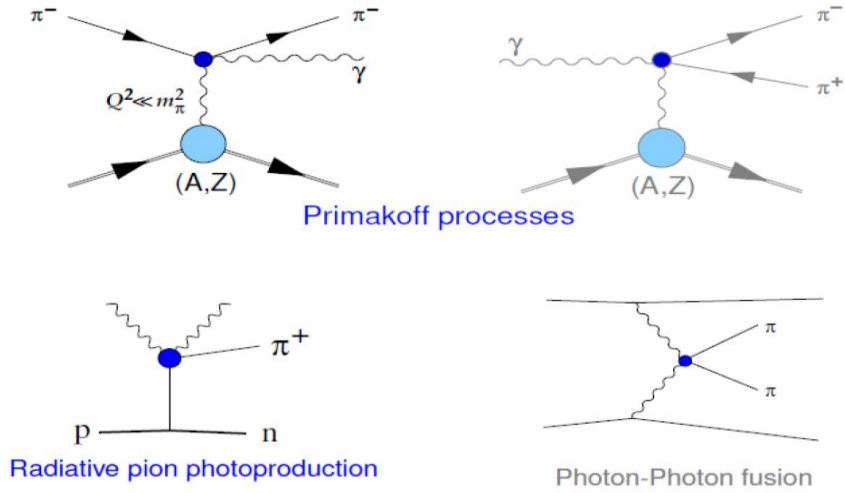

**Fig. 1:** Methods of studying pion polarizabilities: Primakoff Processes: Radiative pion scattering (Bremsstrahlung) on quasi-real photons in the nuclear Coulomb field; Primakoff scattering of high energy γ's in the nuclear Coulomb field leading to two-photon fusion production of pion pairs, radiative pion photoproduction from the proton γp→ γπn; two-photon fusion production of pion pairs γγ→ππ via the $e^+e^- \to e^+e^-\pi^+\pi^-$ reaction.

## 2. COMPASS pion polarizabilities

The COMPASS collaboration at CERN determined $\alpha_\pi - \beta_\pi$ by investigating pion Compton scattering $\gamma\pi \to \gamma\pi$ at center-of-mass energies below 3.5 pion masses [3]. Compton scattering was measured via radiative pion Primakoff scattering (Bremsstrahlung of 190 GeV/c negative pions) in the nuclear Coulomb field of the Ni nucleus: $\pi^-$ Ni → $\pi^-$ Ni γ. Exchanged quasi-real photons are selected by isolating the sharp Coulomb peak observed at lowest four-momentum transfers to the target nucleus, $Q^2 < 0.0015$ GeV$^2$/c$^2$. The resulting data are equivalent to γπ → γπ Compton scattering for laboratory γ's having momenta of order 1 GeV/c incident on a target pion at rest. In the reference frame of this target pion, the cross section is sensitive to ($\alpha_\pi - \beta_\pi$) at backward angles of the scattered γ's. This corresponds to the most forward angles in the laboratory frame for the highest energy Primakoff γ's.

Pion Primakoff scattering at COMPASS is an ultra-peripheral reaction on a virtual photon target. The initial and final state pions are at a distance (impact parameter b) more than 100 fm from the target nucleus, significantly reducing meson exchange and final state interactions. This follows from the extremely small four-momentum transfer $Q_{min}$ to the target nucleus in a Primakoff reaction. For the COMPASS experiment, the minimum momentum transfer $Q_{min}$ ranges from 0.15 to 0.77 MeV/c; and the bulk of the cross section is in the range up to $3Q_{min}$. By the uncertainty principle, the average impact parameter b ~ $\hbar/(2Q_{min})$ is then larger than 100 fm.

COMPASS used a 190 GeV/c beam of negative hadrons (96.8% $\pi^-$, 2.4% $K^-$, 0.8% $\bar{p}$). The COMPASS spectrometer has a silicon tracker to measure precise meson scattering angles, electromagnetic calorimeters for γ detection and for triggering, and Cherenkov threshold detectors for K/π separation [7]. Systematic uncertainties were controlled by many tests, including replacing pions by muons while keeping the same beam momentum. The muon Compton scattering cross section is precisely known, since muons have zero polarizabilities. From a 2009 data sample of 60,000 events, the extracted pion polarizabilities were determined.

Assuming $\alpha_\pi+\beta_\pi=0$, the dependence of the laboratory differential cross section on $x_\gamma=E_\gamma/E_\pi$ is used to determine $\alpha_\pi$, where $x_\gamma$ is the fraction of the beam energy carried by the final state $\gamma$. The variable $x_\gamma$ is related to the $\gamma$ scattering angle for $\gamma\pi \to \gamma\pi$, so that the selected range in $x_\gamma$ corresponds to backward scattering, where the sensitivity to $\alpha_\pi-\beta_\pi$ is largest. Let $\sigma_E(x_\gamma,\alpha_\pi)$ and $\sigma_{MC}(x_\gamma,\alpha_\pi)$ denote the experimental and calculated (via Monte Carlo simulation) laboratory frame differential cross section for a pion (polarizability $\alpha_\pi$) as function of $x_\gamma$; such that $\sigma_{MC}(x_\gamma,\alpha_\pi=0)$ denotes the cross section for a point-like pion having zero polarizability. The $\sigma_E(x_\gamma,\alpha_\pi)$ data are obtained after subtracting backgrounds from the $\pi^-$ Ni $\to \pi^-$ Ni $\gamma$ diffractive channel and the $\pi^-$ Ni $\to \pi^-$ Ni $\pi^0$ diffractive and Primakoff channels. Experimental ratios $R_\pi=\sigma_E(x_\gamma,\alpha_\pi)/\sigma_{MC}(x_\gamma,\alpha_\pi=0)$ are shown in Fig. 2. The polarizability $\alpha_\pi$ and its statistical error are extracted by fitting $R_\pi$ to the theoretical expression:

$$R_\pi = 1 - 72.73 \frac{x_\gamma^2}{1-x_\gamma} \alpha_\pi,$$

where $\alpha_\pi$ is given in units of fm$^3$. The best fit theoretical ratio $R_\pi$ is shown in Fig. 2 as the solid curve [3]. Systematic uncertainties were controlled by measuring $\mu^-$ Ni $\to \mu^-$ Ni $\gamma$ cross sections. The main contribution to the systematic uncertainties comes from the Monte Carlo description of the COMPASS setup. Comparing experimental and theoretical $x_\gamma$ dependences of $R_\pi$ yields: $\alpha_\pi = -\beta_\pi = (2.0\pm0.6_{stat}\pm0.7_{syst})\times10^{-4}$fm$^3$ or equivalently $\alpha_\pi-\beta_\pi = (4.0\pm1.2_{stat}\pm1.4_{syst})\times10^{-4}$fm$^3$. The COMPASS data were corrected for higher-order effects via a one-pion loop low energy expansion in ChPT. The corrections were modest, increasing the extracted polarizability values by $0.6\times10^{-4}$ fm$^3$ [3].

**Fig. 2:** Determination of the pion polarizability by fitting the $x_\gamma$ distribution of the experimental ratios $R_\pi$ (data points) to the theoretical (Monte Carlo) ratio $R_T$ (solid line). (From Ref. 3)

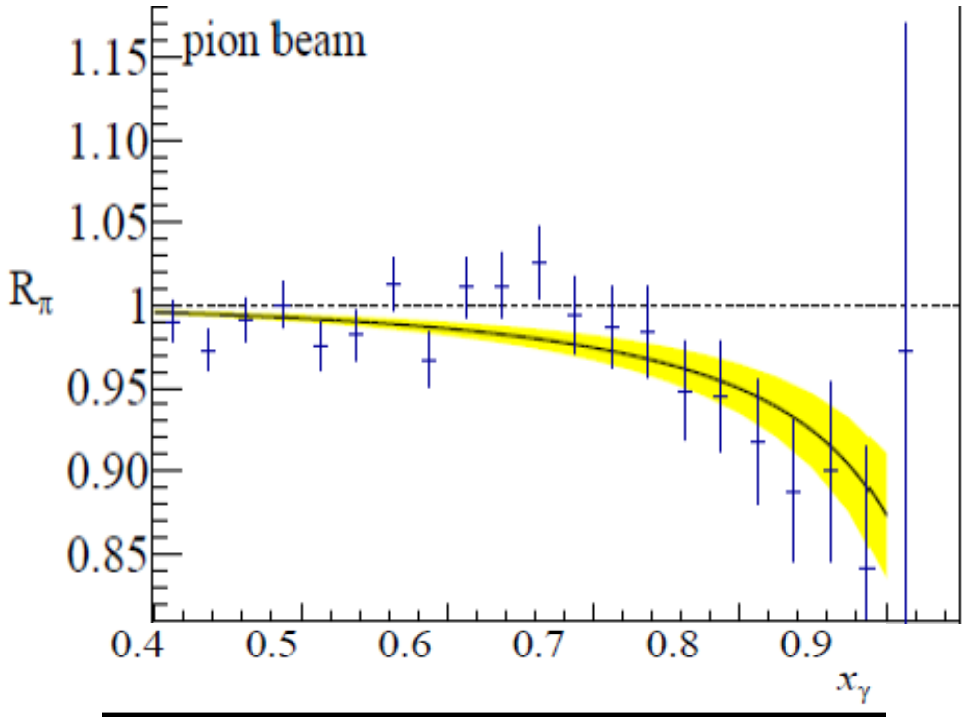

Antipov et al. [8] previously carried out a Primakoff polarizability experiment at Serpukhov using a 40 GeV/c beam of negative pions, and reported $\alpha_\pi - \beta_\pi = 13.6 \pm 2.8_{stat} \pm 2.4_{syst} \times 10^{-4} fm^3$, higher than the COMPASS result. However, since this low statistics experiment (~7000 events) did not allow complete precision studies of systematic errors, their result is not considered further in the present review. Very high statistics data taken by COMPASS in 2012 is expected to provide an independent and high precision determination of $\alpha_\pi$ and $\beta_\pi$ without assuming $\alpha_\pi + \beta_\pi = 0$, and also a first determination of Kaon polarizabilities [1,9].

### 3. MAINZ pion polarizabilities

Radiative $\pi^+$-meson photoproduction from the proton ($\gamma p \rightarrow \gamma \pi^+ n$) was studied at the Mainz Microtron in the kinematic region 537 MeV < $E_\gamma$ < 817 MeV, $140° \leq \theta_{\gamma\gamma'} \leq 180°$, where $\theta_{\gamma\gamma'}$ is the polar angle in the c.m. system of the outgoing gamma and pion [5]. The experimental challenge is that the incident γ-ray is scattered from an off-shell pion, and the polarizability contribution to the Compton cross section from the pion pole diagrams is only a small fraction of the measured cross section. The $\pi^+$-meson polarizability was determined from a comparison of the data with the predictions of two theoretical models, model-1 and model-2 [5,10]. Model-1 includes eleven pion and nucleon pole diagrams by using pseudoscalar coupling; while model-2 includes five nucleon and pion pole diagrams without the anomalous magnetic moments of the nucleons, diagrams for the $\Delta(1232)$, $P_{11}(1440)$, $D_{13}(1520)$, $S_{11}(1535)$ nucleon resonances, and σ-meson contributions. The validity of these two models was studied by comparing the predictions with the experimental data in the kinematic region where the pion polarizability contribution is negligible ($s_1 < 5m_\pi^2$), where $s_1$ is the square of the total energy in the $\gamma\pi^+ \rightarrow \gamma\pi^+$ c.m. system, and where the difference between the predictions of the two models does not exceed 3%. In the region where the pion polarizability contribution is substantial ($5 < s_1/m_\pi^2 < 15$; $-12 < t/m_\pi^2 < -2$), $\alpha_\pi - \beta_\pi$ was determined from a fit of the calculated cross section to the data, as illustrated in Fig. 3. The deduced polarizabilities are $\alpha_\pi - \beta_\pi = (11.6 \pm 1.5_{stat} \pm 3.0_{syst} \pm 0.5_{model}) \times 10^{-4} fm^3$ [5]. The quoted model uncertainty $0.5_{model} \times 10^{-4} fm^3$ denotes the uncertainty associated with using the two chosen theoretical models. It was estimated as half the difference between the model-1 and model-2 polarizability values. However, it does not take into account that comparisons with other possible models may significantly increase the model error. A larger model uncertainty could help explain the difference between COMPASS and Mainz polarizabilities.

It would be of interest to improve the estimate of the model uncertainty by using an independent model to extract the polarizability. A step towards a third model was taken by Kao, Norum, and Wang [11] who studied the $\gamma p \rightarrow \gamma \pi^+ n$ reaction within the framework of heavy baryon chiral perturbation theory. They found that the contributions from two unknown low-energy constants in the πN chiral Lagrangian are comparable with the contributions of the charged pion polarizabilities. Their model therefore gives ~100% uncertainties for the existing Mainz data.

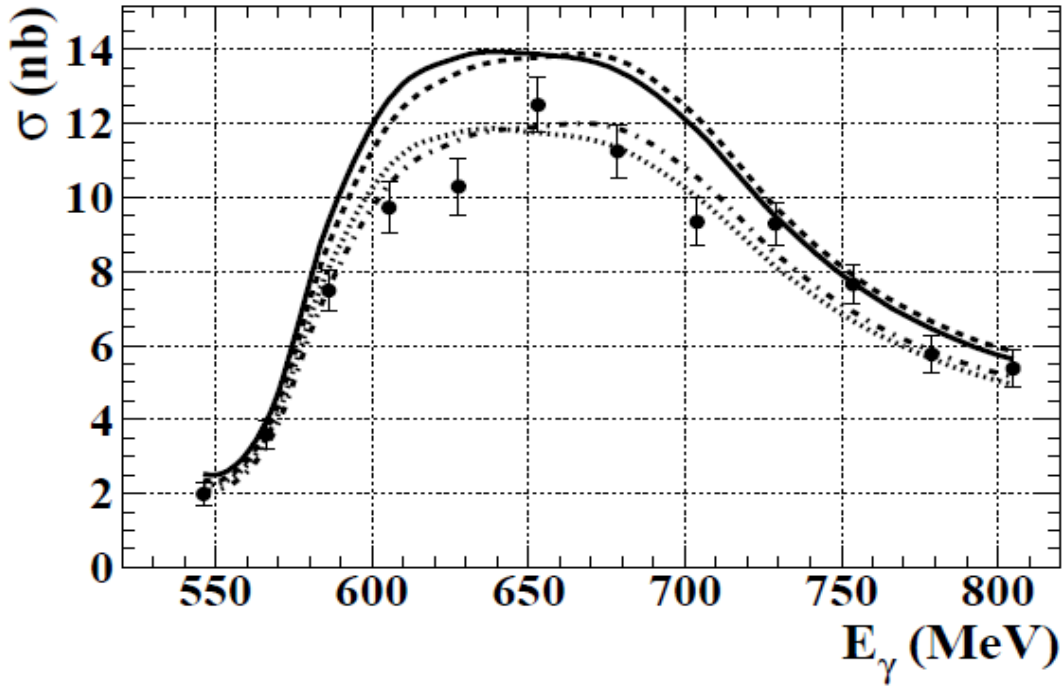

Fig. 3: The cross section of the process $\gamma p \to \gamma \pi^+ n$ integrated over s1 and t in the region where the contribution of the pion polarizability is biggest and the difference between the predictions of the theoretical models under consideration does not exceed 3%. The dashed and dashed-dotted lines are predictions of model-1 and the solid and dotted lines of model-2 for $\alpha_\pi - \beta_\pi = 0$ and $14 \times 10^{-4}$fm³, respectively. From Ref. 5.

### 4. MARK-II pion polarizabilities

Charged pion polarizabilities were determined by comparing MARK-II total cross section data ($\gamma\gamma \to \pi^+\pi^-$) for $M_{\pi\pi} \leq 0.5$ GeV with a ChPT one-loop calculation [12,13]. The MARK-II experiment was carried out via the reaction $e^+e^- \to e^+e^-\pi^+\pi^-$ at a center-of-mass energy of 29 GeV for invariant pion-pair masses $M_{\pi\pi}$ between 350 MeV/c² and 1.6 GeV/c² [4]. Only the region below $M_{\pi\pi} = 0.5$ GeV is considered within the domain of validity of ChPT.

The most important problem in studying the $e^+e^- \to e^+e^-\pi^+\pi^-$ reaction is the elimination of the dominant two-prong QED reactions $e^+e^- \to e^+e^-e^+e^-$ and $e^+e^- \to e^+e^-\mu^+\mu^-$. These leptonic backgrounds below $M_{\pi\pi} = 0.5$ GeV are expected each to be more than 10 times larger than the expected signal. For the critical $M_{\pi\pi}$ region between 350 and 400 MeV/c, MARK-II eliminated these backgrounds by identifying pion pairs using time of flight (TOF), by requiring both tracks to hit an active region of the liquid-argon calorimeter, and by requiring both tracks to have a summed transverse momentum with respect to the $e^+e^-$ axis of less than 150 MeV/c [4]. Summarizing, MARK-II at SLAC has the highest statistics and lowest systematic error data for $\gamma\gamma \to \pi^+\pi^-$ for $M_{\pi\pi} \leq 0.5$ GeV.

A number of theoretical papers subsequently made use of the MARK-II data to deduce pion polarizabilities. For example, theoretical curves from Refs [12,13] are shown in Fig. 4 for Born (dash-dotted line) and ChPT with $\alpha_\pi - \beta_\pi = 5.4 \times 10^{-4}$fm³ (full line). The cross section excess below $M_{\pi\pi} = 0.5$ GeV compared to the Born calculation was interpreted as due to pion polarizabilities, with best fit value $\alpha_\pi - \beta_\pi = (4.4 \pm 3.2_{stat+syst}) \times 10^{-4}$ fm. Similar analyses from Refs. [14,15] gave $\alpha_\pi - \beta_\pi \sim 5.3 \times 10^{-4}$fm³, consistent with this result.

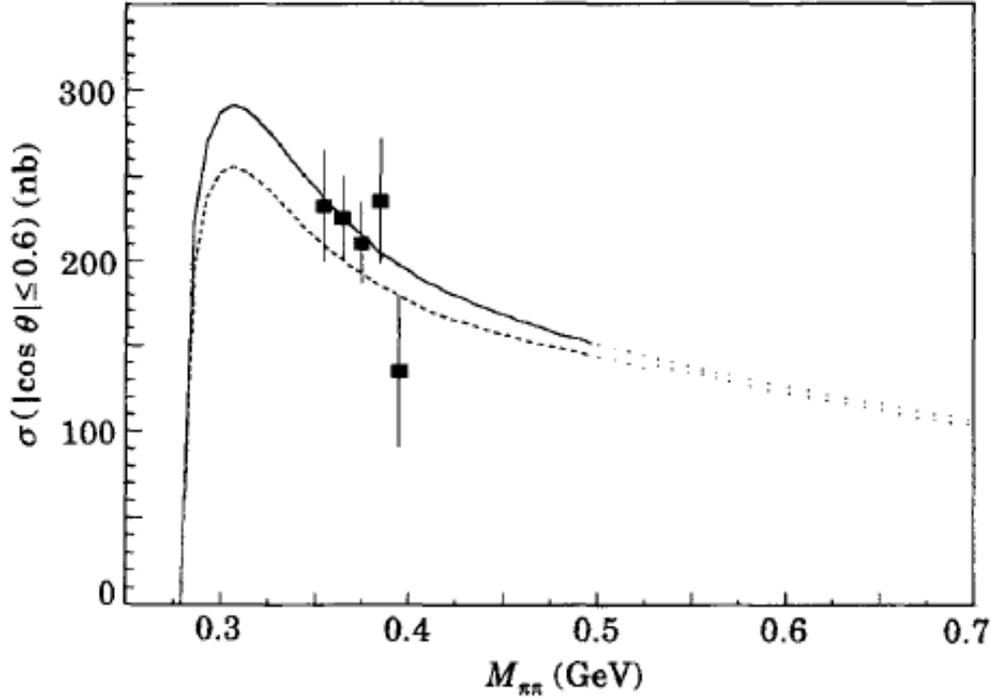

Fig. 4: MARK-II total cross section data ($\gamma\gamma \to \pi^+\pi^-$) for $M_{\pi\pi} \leq 0.5$ GeV. The theoretical curves are: Born (dash-dotted line); ChPT with $\alpha_\pi - \beta_\pi = 5.4 \times 10^{-4}$ fm$^3$ (full line). The region above $M_{\pi\pi} = 0.5$ GeV is considered outside the domain of validity of ChPT. From Ref. 4.

### 5. Dispersion Relations and Pion Polarizabilities

Pion polarizabilities are determined by how the $\gamma\pi \to \gamma\pi$ Compton scattering amplitudes approach threshold. By crossing symmetry, the $\gamma\pi \to \gamma\pi$ amplitudes are related to the $\gamma\gamma \to \pi\pi$ amplitudes. Dispersion relations (DRs) provide the method to continue the $\gamma\gamma$ amplitudes analytically to the Compton scattering threshold. DRs describe how pion polarizabilities contribute to both $\gamma\gamma \to \pi^+\pi^-$ and $\gamma\gamma \to \pi^0\pi^0$ reactions [16,17].

Most recently, Dai and Pennington (DP) carried out DR calculations [18]. In their formalism, the $\pi^0$ and $\pi^\pm$ polarizability values are correlated, so that knowing one allows calculating the other. Their calculation provides important guidance to the planned JLab $\gamma\gamma \to \pi^+\pi^-$ experiment [6], that the cross section must be measured to better than 2.2 nb to fix $\alpha_\pi - \beta_\pi$ to an accuracy of 10%. Using COMPASS $\alpha_\pi - \beta_\pi = 4.0 \times 10^{-4}$ fm$^3$ and Mainz $\alpha_\pi - \beta_\pi = 11.6 \times 10^{-4}$ fm$^3$ as input, DP calculate both $\gamma\gamma \to \pi^+\pi^-$ and $\gamma\gamma \to \pi^0\pi^0$ cross sections. They compare these with MARK-II $\gamma\gamma \to \pi^+\pi^-$ data and DESY Crystal Ball $\gamma\gamma \to \pi^0\pi^0$ data [19]. With the COMPASS value, they find excellent agreement for $\gamma\gamma \to \pi^+\pi^-$ and reasonable agreement for $\gamma\gamma \to \pi^0\pi^0$. With the Mainz value, their DR calculations and Crystal Ball data do not agree at all. The differences are too large to be explained by uncertainties in the DP calculation. DP conclude that $\alpha_\pi - \beta_\pi = 11.6 \times 10^{-4}$ fm$^3$ is excluded by the Crystal Ball $\gamma\gamma \to \pi^0\pi^0$ data. Following their evaluation, and also considering the large difference between Mainz and COMPASS-MARK-II values, the Mainz polarizability value is excluded from the Section 6 summary discussion.

Corrections for higher-order effects via a one-pion loop low energy expansion in ChPT increased the COMPASS polarizability values by $0.6\times10^{-4}$ fm$^3$ [3]. Pasquini showed using subtracted Dispersion Relations (DR) for the pion Compton amplitude [9,20] that the yet higher order energy contributions neglected in the COMPASS analysis are very small. DRs take into account the full energy dependence, while ChPT uses a low energy expansion. In the COMPASS kinematic region of interest, the ChPT one-loop calculation and subtracted DRs agree in the mass range up to $4m_\pi$ at the two per mille level [9,20]. Furthermore, the DR predictions of Pasquini, Drechsel, and Scherer [21,22], using unsubtracted DRs for the $\gamma\gamma\to\pi\pi$ amplitude, agree with the results of ChPT.

By contrast, Filkov & Kashevarov (FK) claim that there is significant disagreement between the ChPT one-loop calculation and their DR calculation [23,24]. They claim therefore that the higher order corrections cannot be made via a one-pion loop calculation in ChPT. In their DR calculation, the contribution of the σ-meson to the COMPASS pion Compton scattering cross section is very substantial [23,24]. FK claim that the COMPASS deduced $(\alpha_\pi-\beta_\pi)$ is very sensitive to σ-meson contributions. They find via their DR calculations, taking into account the contribution of the σ-meson, that $(\alpha_\pi-\beta_\pi) \sim 11\times10^{-4}$fm$^3$ for the COMPASS experiment.

Pasquini, Drechsel, and Scherer (PDS) claim however that the FK discrepancies arise due to the way that they implement the dispersion relations [21,22]. DRs may be based on specific forms for the absorptive part of the Compton amplitudes. The scalar σ-meson has low mass ($M_\sigma \sim 441$ MeV) and large width ($\Gamma_\sigma \sim 554$ MeV). In the DR calculation of FK, this resonance is modelled by an amplitude characterized by a pole, width and coupling constant. They choose an analytic form (small-width approximation, and an energy-dependent coupling constant) that leads to a dispersion integral that diverges like $1/\sqrt{t}$ for $t \to 0$. PDS examined the analytic properties of different analytic forms, and showed that the strong enhancement by the σ-meson, as found by FK, is connected with spurious (unphysical) singularities of this nonanalytic function. PDS explain that the FK resonance model gives large (unstable) results for the pion polarizability, because it diverges at $t = 0$, precisely where the polarizability is determined. That is, via DRs, the imaginary part of the Compton amplitudes serves as input to determine the polarizabilities at the Compton threshold ($s = m_\pi^2$, $t = 0$). PDS found that if the basic requirements of dispersion relations are taken into account, DR results and effective field theory are consistent. FK disagree with these PDS conclusions, claiming that their DR results are not due to spurious singularities. The present status report follows the views of Pasquini, Drechsel, and Scherer. Besides the σ-meson, modifications of COMPASS data analysis suggested by Filkov and Kashevarov [23,24] were studied, and found not to affect the deduced polarizabilities [3].

## 6. Comparison of pion polarizability data with ChPT

The pion is believed to belong to the pseudoscalar meson nonet and to be one of the Goldstone bosons associated with spontaneously broken chiral symmetry. Chiral perturbation theory (ChPT) is therefore expected to successfully describe the electromagnetic interactions of pions. In this framework, the low-energy interactions of the pion are described by a phenomenological effective Lagrangian which stems directly from QCD, with only the assumptions of chiral symmetry $SU(3)_L \times SU(3)_R$, Lorentz invariance and low momentum transfer. Unitarity is achieved by adding pion loop corrections to lowest order, and the resulting infinite divergences are absorbed into physical (renormalized) coupling constants $L_i^r$ (tree level coefficients in the Lagrangian $L^{(4)}$, see ref. [25,26]). In particular, with a perturbative expansion of $L^{(4)}$, limited to terms quartic in the external momenta and pion mass $O(p^4)$, the method establishes relationships between different processes in terms of a common set of renormalized parameters $L_i^r$. At $O(p^4)$ level, the perturbative expansion is truncated at terms quartic in the photon momentum and 12 coupling constants are needed. For example, in the charged pion case, the ratio of axial to vector form factor from radiative pion beta-decay and the electric polarizability are expressed as $h_A/h_V = 32\pi^2(L_9^r + L_{10}^r)$ and $\alpha_\pi = -\beta_\pi = 4\alpha (L_9^r + L_{10}^r)/m_\pi F_\pi^2$, where $F_\pi$ is the pion decay constant and $\alpha$ is the fine-structure constant [25,26]. The electric $\alpha_\pi$ and magnetic $\beta_\pi$ charged pion Compton polarizabilities are therefore of fundamental interest in the low-energy sector of quantum chromodynamics (QCD). From the above description, they are directly linked to the phenomenon of spontaneously broken chiral symmetry within QCD and to the dynamics of the pion-photon interaction. The experimental pion beta-decay ratio $h_A/h_V = 0.47 \pm 0.03$ [27] then leads to the one-loop prediction $\alpha_\pi - \beta_\pi = 5.6 \pm 0.4 \times 10^{-4} \text{fm}^3$, assuming $\alpha_\pi + \beta_\pi = 0$. The ChPT two-loop prediction is $\alpha_\pi - \beta_\pi = (5.7 \pm 1.0) \times 10^{-4} \text{ fm}^3$ and $\alpha_\pi + \beta_\pi = 0.16 \times 10^{-4} \text{ fm}^3$ [2]. The present report focuses on ChPT, and reviews all the different available pion polarizability experiments. Not within the framework of ChPT, Holstein showed that meson exchange via a pole diagram involving the $a_1(1260)$ resonance provides the main contribution ($\alpha_\pi - \beta_\pi = 5.2 \times 10^{-4} \text{fm}^3$) to the polarizability [28]. Other recent polarizability reviews are also available [9,29,30,31,32].

Pion polarizabilities affect the shape of the $\gamma\pi$ Compton scattering angular distribution. The pion polarizability combination ($\alpha_\pi - \beta_\pi$) was measured by: (1) CERN COMPASS via radiative pion Primakoff scattering (pion Bremsstrahlung) in the nuclear Coulomb field, $\pi Z \to \pi Z \gamma$, equivalent to $\gamma\pi \to \gamma\pi$ Compton scattering for laboratory $\gamma$'s having momenta of order 1 GeV/c incident on a target pion at rest; and (2) SLAC PEP Mark-II via two-photon production of pion pairs, $\gamma\gamma \to \pi^+\pi^-$. Only the most recent measurements by these two methods are considered here, as they have the significantly highest statistics and smallest systematic errors:
(1) $\alpha_\pi - \beta_\pi = (4.0 \pm 1.2_{stat} \pm 1.4_{syst}) \times 10^{-4} \text{fm}^3$, (2) $\alpha_\pi - \beta_\pi = (4.4 \pm 3.2_{stat+syst}) \times 10^{-4} \text{ fm}^3$. These polarizability values are in good agreement with the two-loop ChPT prediction $\alpha_\pi - \beta_\pi = (5.7 \pm 1.0) \times 10^{-4} \text{ fm}^3$, thereby strengthening the identification of the pion with the Goldstone boson of QCD.